\documentclass[a4paper,11pt]{article}
\pdfoutput=1 

\usepackage{jheppub} 

\usepackage[T1]{fontenc} 
\usepackage{booktabs}
\usepackage{tabularx}
\usepackage{subfig}
\usepackage{latexsym}
\usepackage{slashed}
\usepackage{snapshot}
\usepackage{natbib}

\preprint{\parbox{5cm}{CERN-PH-TH-2014-261}}
\renewcommand{\arraystretch}{1.3}

\title{How-to: Write a parton-level Monte Carlo particle physics event generator}

\author{Andreas Papaefstathiou,}

\affiliation{PH Department, TH Department, CERN, CH-1211 Geneva 23, Switzerland \& Higgs Centre for Theoretical Physics, University of Edinburgh, Peter Guthrie Tait Road, Edinburgh EH9 3FD, UK.}
\emailAdd{apapaefs@cern.ch}

\abstract{This article provides an introduction to the principles of particle physics event generators that are based on the Monte Carlo method. Following some preliminaries, instructions on how to build a basic parton-level Monte Carlo event generator for the hard interaction are given through exercises. Indications on how to proceed to full event simulations are given.\footnote{The related course was given as part of the ``Advanced Scientific Computing Workshop'' at ETH Z\"urich in July 2014.}}

\begin{document} 
\maketitle
\flushbottom

\section*{Preface}
According to Wikipedia, ``a how-to is an informal, often short, description of how to accomplish a specific task. A how-to is usually meant to help non-experts, may leave out details that are only important to experts, and may also be greatly simplified from an overall discussion of the topic.''~\cite{wikipediahowto}. In some aspects this is also valid for this article. However, in this case the aim of this article is to provide some insight to the experts themselves, that is, physicists, who may use Monte Carlo event generators as ``black boxes'' to serve their purposes, either to calculate cross sections or to generate events to further simulate and investigate a future possible experimental analysis. 
\addcontentsline{toc}{section}{Preface}

\section{Introduction}
Treatment of particle collisions in mechanics starts off relatively easy: we initially study elastic collisions of two spheres in one dimension, and we are asked to calculate the various momenta after a collision occurs. The next complication involves adding the effects of inelasticity. This results in some energy loss e.g. through the balls sticking together and so on. The theoretical description of collisions of elementary particles starts off equally simply: the scattering of two electrons, for example, can be simulated at the first order in the perturbative picture (leading order), via the exchange of a single photon, representing an elastic collision. However, ``Truth is stranger than Fiction, but it is because Fiction is obliged to stick to possibilities; Truth isn't.''~\cite{twain}. In the context of particle physics, to describe `Truth', i.e. Nature, in our `fictional' simulations, we need to model a multitude of effects using a series of approximations and models. To name but a few of these aspects:
\begin{itemize}
\item particles radiate, e.g. photons off electrons, gluons off quarks,
\item incoming particles may be confined in a bound state, e.g. quarks and gluons in protons, 
\item higher-order corrections in perturbation theory are too laborious to compute beyond the first few orders, 
\item the phase space of the final-state particles is huge and of a variable number of dimensions,
\item and many effects cannot be described by perturbation theory and need to be modelled. 
\end{itemize}

Many of the above effects have been incorporated into computer simulations using Monte Carlo techniques. 

The large dimensionality of the phase space makes the Monte Carlo integration the method of choice. The Markovian nature of the parton shower process can also be formulated as a Monte Carlo process. For different aspects of the simulation, several tools already exist on the ``market''. These serve many purposes, sometimes overlapping, following different approaches and methodologies. Without (and far from) being completely inclusive, some of these tools are (i) \texttt{MadGraph}, (ii) \texttt{HERWIG 7}, (iii) \texttt{Pythia 8} and (iv) \texttt{Sherpa}. \texttt{MadGraph} provides parton-level events of automatically generated process that the user asks for, at the moment capable of generating events at leading order and next-to-leading order in QCD (via the MC@NLO method). The output can then be given to a general-purpose event generator for showering and hadronization~\cite{Alwall:2011uj, Alwall:2014hca}. \texttt{HERWIG 7}~\cite{Bahr:2008pv, Arnold:2012fq, Gieseke:2011na, Bellm:2019zci, Bellm:2017bvx, Bellm:2015jjp, Bellm:2013hwb}\footnote{Formerly known as \texttt{HERWIG++}.}, \texttt{Pythia 8}~\cite{Sjostrand:2006za,Sjostrand:2007gs} and \texttt{Sherpa}~\cite{Gleisberg:2008ta} are general-purpose event generators that include in part some automation for generating processes at parton level as well as taking into account the effects of the parton shower, hadronization and the underlying event.

For a review of the detailed physics and the philosophy behind Monte Carlo event generators, I refer the reader to Ref.~\cite{Buckley:2011ms}. Here we wish to examine the minimal aspects of constructing a parton-level event generator, adding some hints at the end for how one can incorporate the more advanced features such as a parton shower, hadronization, the underlying event and including higher-order corrections. We will start with some preliminaries in the next section.
\section{Preliminaries}
\subsection{Monte Carlo integration}\label{sec:mcint}
This section has been adapted in part from Peter Richardson's CTEQ 2006 lectures\footnote{\url{http://www.ippp.dur.ac.uk/~richardn/talks/}.} as well as Mike Seymour's PhD thesis, Chapter 3.\footnote{\url{http://www.hep.manchester.ac.uk/u/seymour/thesis/}.} 

Monte Carlo integration is based on a simple observation: the value of an integral can be recast as the average of the integrand:
\begin{equation}
I = \int_{x_1}^{x_2} \mathrm{d}x~ f(x) = (x_2 - x_1 ) \left< f(x) \right> \;.
\end{equation}
Consequently, this implies that if we take some, say $N$, values of $x$, distributed uniformly in $(x_1, x_2)$, then the average of $f(x)$ will be a good estimator of the integral, $I$. We can then write:
\begin{equation}
I \approx (x_2 - x_1) \frac{1}{N} \sum_{i=1}^N f(x_i) \;.
\end{equation}
We can draw the values $x_i$ randomly: if $\rho_i$ is a uniform random number in $(0,1)$,\footnote{That is, with equal probability to lie anywhere within the given interval.} then we have:
\begin{equation}
x_i = (x_2 - x_1 ) \rho_i + x_1 \;.
\end{equation}
To estimate the accuracy of the calculation we can employ the Central Limit Theorem: the distribution of $\left< f(x) \right>$ will tend to a Gaussian with standard deviation $\sigma_\mathrm{MC} = \sigma / \sqrt{N}$, where $\sigma$ is the standard deviation of $f(x_i)$. Our inaccuracy simply decreases as $1/\sqrt{N}$. We often also define the weight: $W_i = (x_2 - x_1) f(x_i)$, and then the integral is simply the average of the weight:
\begin{equation} \label{eq:integral}
I \approx I_N = \frac{1}{N} \sum_{i=1}^N W_i \;.
\end{equation}
We also define the variance, $V_N \equiv \sigma^2$:
\begin{equation}
V_N = \frac{1}{N} \sum_i W_i^2 - \left[ \frac{1}{N} \sum_i W_i \right]^2 \;,
\end{equation}\label{eq:mcerror}
from which $\sigma_\mathrm{MC} = \sqrt{ V_N / N }$, and we finally arrive at the expression:
\begin{equation}
I \approx I_N \pm \sqrt{ \frac{V_N} { N } } \;.
\end{equation}
One can compare the convergence of the Monte Carlo integration technique to those for other common techniques. In $d$-dimensions the convergence of techniques such as the `Trapezium Rule', `Simpson's rule' and Gaussian quadrature are $\propto 1/N^{2/d}$, $\propto 1/N^{4/d}$ and $1/N^{(2m-1)/d}$ respectively. On the other hand, Monte Carlo integration always extends trivially and converges as $\propto 1/\sqrt{N}$ in $d$ dimensions, and hence converges already faster than all the aforementioned methods in $d>4$, $d>8$ and $d>4m-2$ respectively.\footnote{A summary of the rate of convergence of the various techniques is given in Table~\ref{tb:convergence} in Appendix~\ref{app:convergence}.} In typical LHC events we have $\mathcal{O}(1000)$ particles and hence this results in $O(3000)$ phase space integrals. Therefore, Monte Carlo integration is in fact the only viable option. 

The biggest disadvantage of the Monte Carlo method is the relatively slow divergence in few dimensions. This can be tackled by `Importance Sampling', which we will discuss below. Its principal advantages over numerical quadrature can be summarised as:
\begin{itemize}
\item fast convergence in many dimensions,
\item arbitrarily complex integration regions,
\item small feasibility limit: the minimum number of functional evaluations which must be made for the method to work at all, in this case 2,
\item small growth rate: the smallest number of additional function evaluations needed to improve the current estimate, in this case 1: each additional point improves the estimate of the integral,
\item easy estimate of accuracy.
\end{itemize}
\subsection{Improving convergence of the Monte Carlo integration}\label{sec:jacob}
The accuracy of an integral calculated via the Monte Carlo integration method is given by $ \sqrt{ V_N  / N  } $. Thus one can simply increase the number of points to increase the accuracy. However, one can also look for ways to decrease $V_N$, e.g., by a method called `Importance Sampling'~\cite{Kittel:1970rh}. The basic idea is to perform a Jacobian transform so that the integral is flatter in the new integration variable. This is equivalent to finding a transform such that $V_N' < V_N$. 

\begin{figure}[!htb]
\centering
\includegraphics[width=0.75\columnwidth]{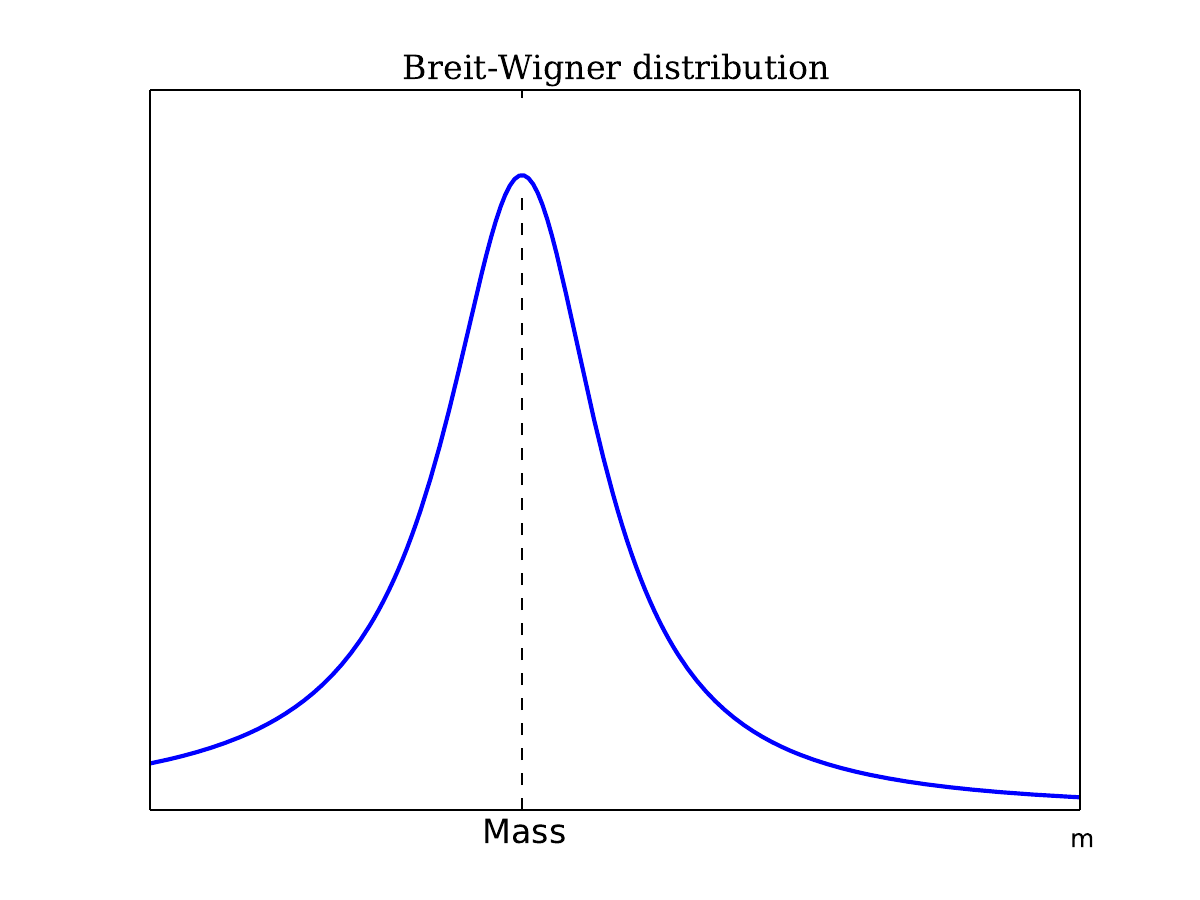}
\caption{An example of the Breit-Wigner distribution, made for $M=90$, $\Gamma=10$.}
\label{fig:bw}
\end{figure}

We begin by considering the simplest case encountered in particle physics. In cross section calculations we often encounter the so-called Breit-Wigner distribution, that describes the `peak' of a resonance:
\begin{equation}
F_\mathrm{BW}(m^2) = \frac{1}{ (m^2 - M^2)^2 + M^2 \Gamma^2 } \;,
\end{equation} 
where $M$ would be the physical (on-shell) mass of the particle, $m$ is the off-shell mass and $\Gamma$ its width. An example of the distribution (made using $M=90$, $\Gamma = 10$) is shown in Fig.~\ref{fig:bw}.

We then often encounter integrals of the form:
\begin{equation}\label{eq:bwint}
I = \int_{M_\mathrm{min}^2}^{M_\mathrm{max}^2} \mathrm{d} m^2 \frac{1}{ (m^2 - M^2)^2 + M^2 \Gamma^2 } \;.
\end{equation}
The transformation we wish to consider is $m^2 \rightarrow \rho$, where
\begin{equation}
m^2 = M\Gamma \tan \rho + M^2 \;,
\end{equation}
and the corresponding Jacobian is given by:
\begin{equation}
J = \left| \frac{ \partial m^2 } { \partial \rho }\right| = M\Gamma \sec ^2 \rho\;.
\end{equation}
Hence we have: 
\begin{eqnarray}
I  &=&  \int_{\rho_\mathrm{min}}^{\rho_{max}} \mathrm{d} \rho \left| \frac{ \partial m^2 } { \partial \rho }\right| \frac{1}{ (m^2 - M^2)^2 + M^2 \Gamma^2 } \nonumber\\
  &=& \frac{1}{M\Gamma} \int_{\rho_\mathrm{min}}^{\rho_{max}} \mathrm{d} \rho \;.
\end{eqnarray}

It is evident that in this case, we have in fact reduced the variance to zero: $V_N' = 0$. In practice, few of the cases we need to deal with can be exactly integrated. In cases of complicated integration regions, one can try and pick a function that approximates the behaviour of the function we want to integrate. A specific method, called multi-channel integration, aims to handle the situation where one is faced with multiple peaks in the phase space and one can then not use a single Breit-Wigner. The method can be automated and is used in all modern Monte Carlo event generators~\cite{Lepage:1980dq}.

\subsection{Hit-or-Miss Monte Carlo}\label{sec:hitmiss}
There are two main aspects of Monte Carlo that make it ideal for use in constructing event generators: the close relationship between the numerical method and the physical process under study, both being `random' in some sense, and the ability to make unweighted events. 

In a similar way that a Monte Carlo integration of the sort described in Section~\ref{sec:mcint} is performed, one can scan over the function $f(x)$ and collect a set of phase-space points, along with their associated probabilities, corresponding to the weight of each in the integral. These points effectively correspond to possible `events', with their weights corresponding to their probability of occurring. However, if we want to use these events, e.g. to perform an experimental analysis, then we must always carry the associated weight around for use in histograms, averages and so on. This can be inconvenient but also very inefficient: time may be wasted in some latter part of the simulation (e.g. detector simulation) to events that possess only a very small weight. The so-called `hit-or-miss' method aims to equalize the weights of different events as far as possible. 

Since the weight of each event is proportional to the probability of it occurring, we can unweigh the events by keeping only a fraction of them, according to their weights. We do this by finding the maximum weight which occurs in the integration region. This can be done while performing Monte Carlo integration. We choose to keep (`accept') each event with probability $f(x)/f_\mathrm{max}$. The rest are thrown away (`rejected'). All accepted events are given a weight $\left< f \right>$, calculated from the Monte Carlo integral over all generated events (not just the accepted events). The complete algorithm for integration and event generation is then: 
\begin{enumerate}
\item Monte Carlo integration and scanning are performed: $N$ points are picked randomly, according to some distribution and their weight is accumulated to the sums: $\sum_i W_i$, $\sum_i W_i^2$. The cross section and corresponding error are computed according to Eqs.~\ref{eq:integral} and~\ref{eq:mcerror}. During this period, the phase-space point which give the maximum weight, $W_\mathrm{max}$ is stored. 
\item Generating unweighted events via the `hit-or-miss' method: go through randomly chosen phase-space points and compare the probability of each, given by $W_i/W_\mathrm{max}$ to a random number $R \in (0,1)$. If $W_i/W_\mathrm{max} > R$, we `accept' the event, otherwise we reject it. This is done until we have collected the desired number of events, $N_\mathrm{events}$. 
\end{enumerate}
\subsection{Factorisation and the structure of event generators}
\begin{figure}[!htb]
  \centering
    \includegraphics[width=0.75\linewidth]{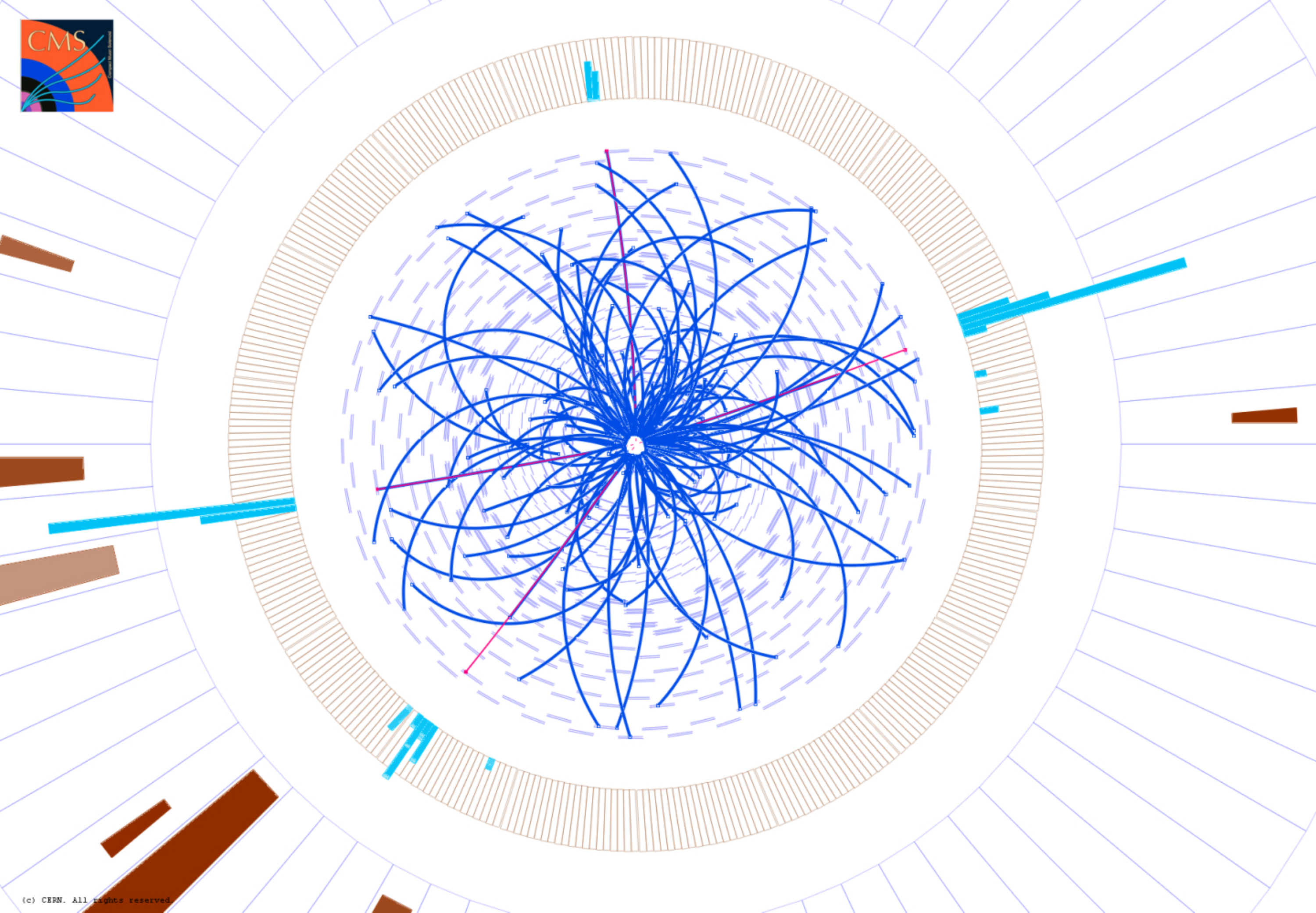}
  \caption{Real CMS proton-proton collision events in which four high energy electrons are observed. The event shows characteristics expected from the decay of a Higgs boson but is also consistent with background Standard Model physics processes.}
  \label{fig:cmsevent}
\end{figure}

The complexity of an event is something that we (particle physicists) are all familiar with. This is exemplified in Fig.~\ref{fig:cmsevent}. Even if the hard collision is simple, we expect thousands of final state particles at hadron colliders. It is evident that this poses many challenges in simulating events: it is difficult or even impossible to construct an efficient algorithm but also hard to exactly calculate final-state distributions of hadrons. 

\begin{figure}[!htb]
  \centering 
  \includegraphics[scale=0.55, angle=0]{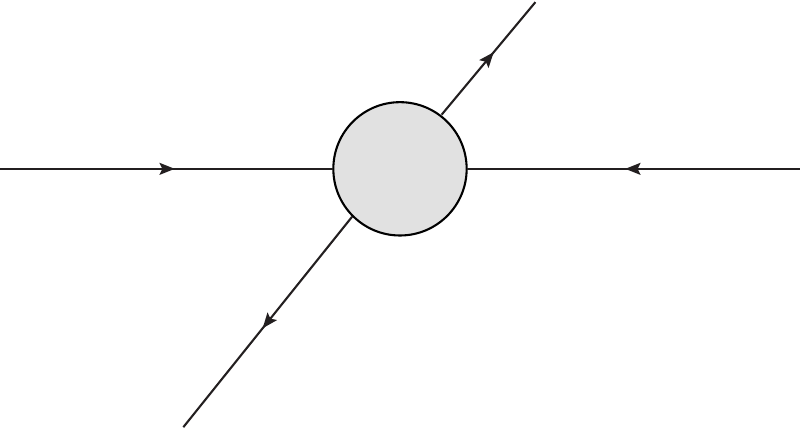}
  \caption[]{\textbf{STEP 1}: Generation of the hard process.}
\label{fig:step1}
\end{figure}

\begin{figure}[!htb]
  \centering 
  \includegraphics[scale=0.55, angle=0]{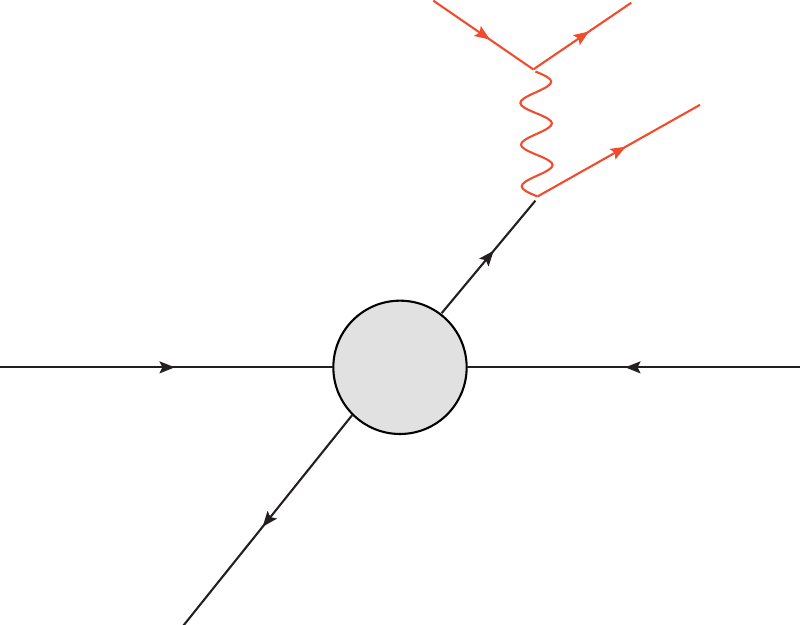}
  \caption[]{\textbf{STEP 2}: Decay of heavy resonances.}
\label{fig:step2}
\end{figure}

\begin{figure}[!htb]
  \centering 
  \includegraphics[scale=0.55, angle=0]{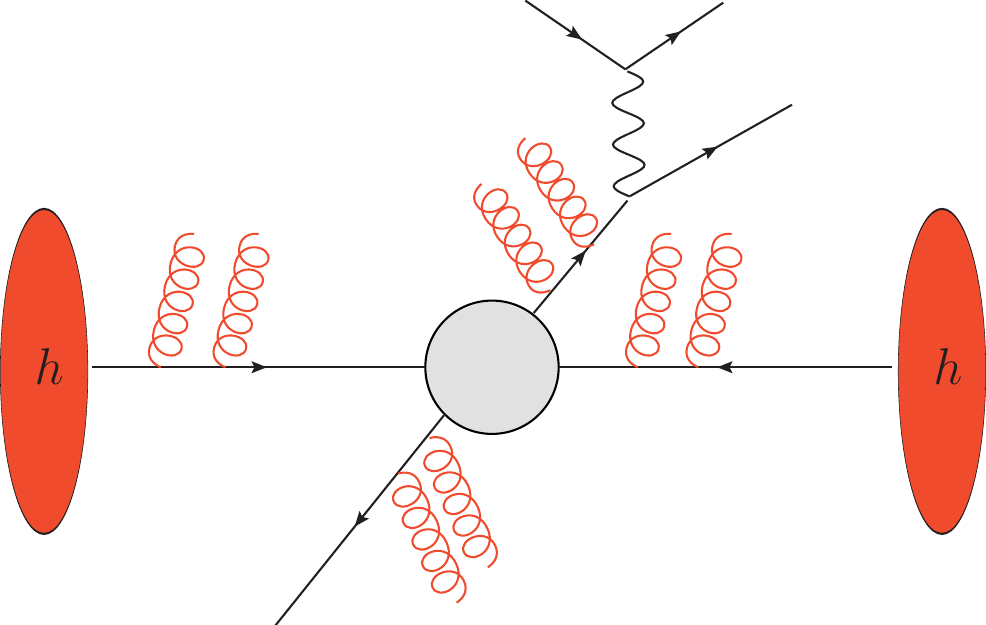}
  \caption[]{\textbf{STEP 3}: Parton showers.}
\label{fig:step3}
\end{figure}

\begin{figure}[!htb]
  \centering 
  \includegraphics[scale=0.55, angle=0]{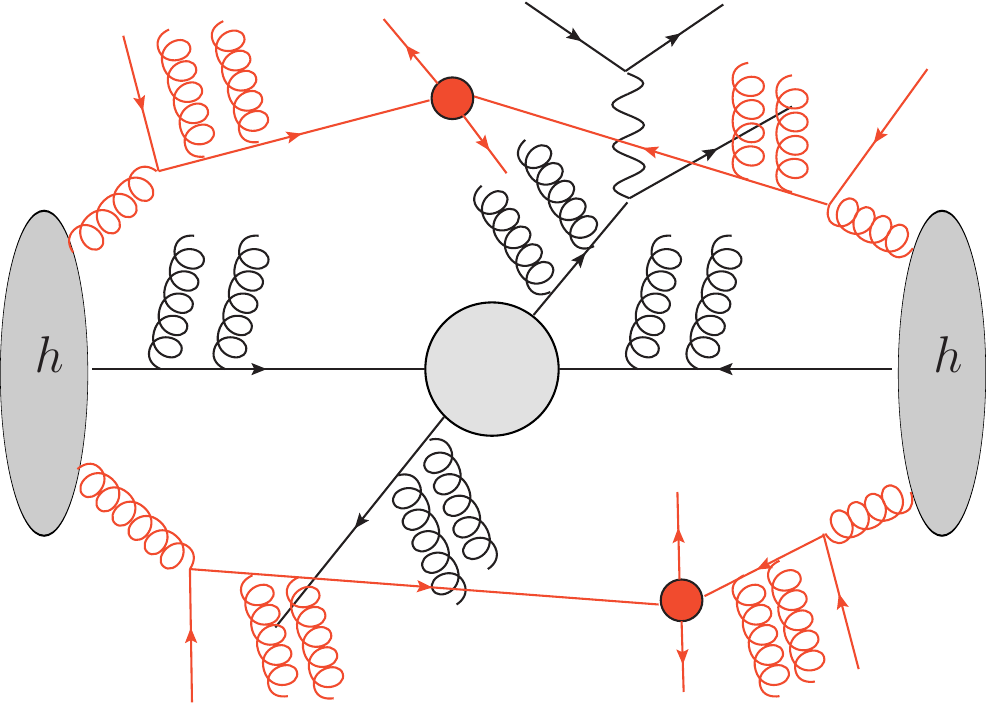}
  \caption[]{\textbf{STEP 4}: Multiple parton interactions.}
\label{fig:step4}
\end{figure}

\begin{figure}[!htb]
  \centering 
  \includegraphics[scale=0.55, angle=0]{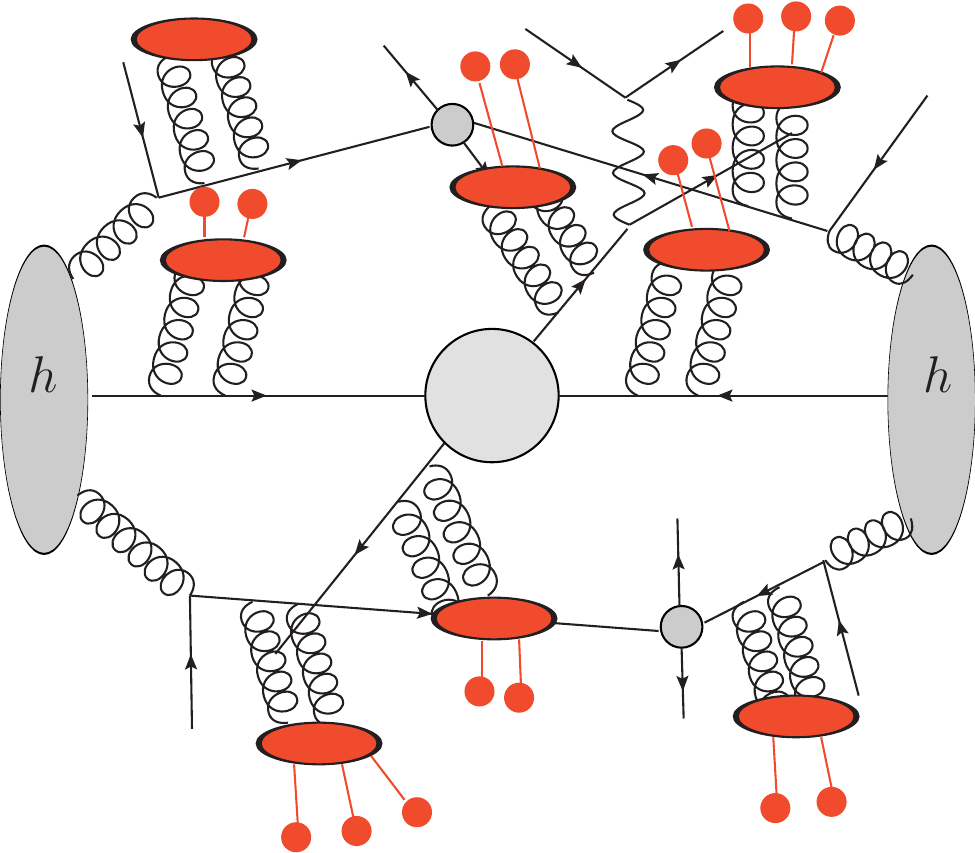}
  \caption[]{\textbf{STEP 4}: Hadronization and hadron decays.}
\label{fig:step5}
\end{figure}

It is fortunate that the probabilities for separate stages of the events factorize in some well-motivated approximations. This is akin to the ``adiabatic approximation'', where e.g. if the support of a rigid pendulum is moving at a frequency much lower than the natural frequency of the pendulum, the two motions can be treated independently or in a ``factorised'' way. We will not examine these stages in detail here: instead, we illustrate a possible, and common, factorisation of an event with the help of schematic diagrams as performed by a generic event generator when producing full event simulation. Figs.~\ref{fig:step1} to~\ref{fig:step5} demonstrate the various steps~\cite{Papaefstathiou:2011rc}. In each step, the newly appearing features are highlighted in red. In the present article we will only examine how step 1 is implemented in a numerical simulation.

\begin{enumerate}
\item{\textbf{Hard process generation,
      Figure~\ref{fig:step1}}: The hard process is
    generated by choosing a point on the phase space according to the
    `hit-or-miss' method.}
\item{\textbf{Heavy resonance decay,
      Figure~\ref{fig:step2}}: Heavy resonances
    with narrow widths are
    decayed before the parton shower. In this example the heavy
    resonance could be a top quark, decaying to a $\ell \nu_{\ell}$
    and a $b$-quark.} 
\item{\textbf{Parton showers,
      Figure~\ref{fig:step3}}: The incoming partons
    are showered by ``evolving backwards'' to the incoming hadrons,
    producing initial-state radiation. Any final-state particles
    that are colour-charged also radiate, producing final-state
    radiation.}
\item{\textbf{Multiple parton interactions,
      Figure~\ref{fig:step4}}: Secondary, lower-energy
    interactions between partons within the colliding hadrons,
    modelled as QCD $2\rightarrow2$ interactions, are generated.}
\item{\textbf{Hadronization and hadron decays,
      Figure~\ref{fig:step5}}: In the cluster
    model of hadronization, clusters of coloured (QCD-charged) particles are formed and hadrons are produced. Unstable
    hadrons are subsequently decayed.}
\end{enumerate}

\label{app:mcillustration}

\section{Exercises}
The exercises and solutions can be found at: \url{https://apapaefs.web.cern.ch/apapaefs/mchowto.html} and are also attached to this document's source. We first review the necessary particle physics input and consider two exercises: the Monte Carlo simulation of the $e^+e^- \rightarrow  \gamma \rightarrow \mu^+ \mu^-$ process at lepton colliders and of $q\bar{q} \rightarrow Z/\gamma \rightarrow \mu^+ \mu^-$ at hadron colliders. 
\subsection{Particle physics input}
We first provide some basic formulae that we will employ in the exercises given in this section. 
\subsubsection{$e^+e^- \rightarrow  \gamma \rightarrow \mu^+ \mu^-$} 
\begin{figure}[!htb]
  \centering
    \includegraphics[width=0.5\linewidth]{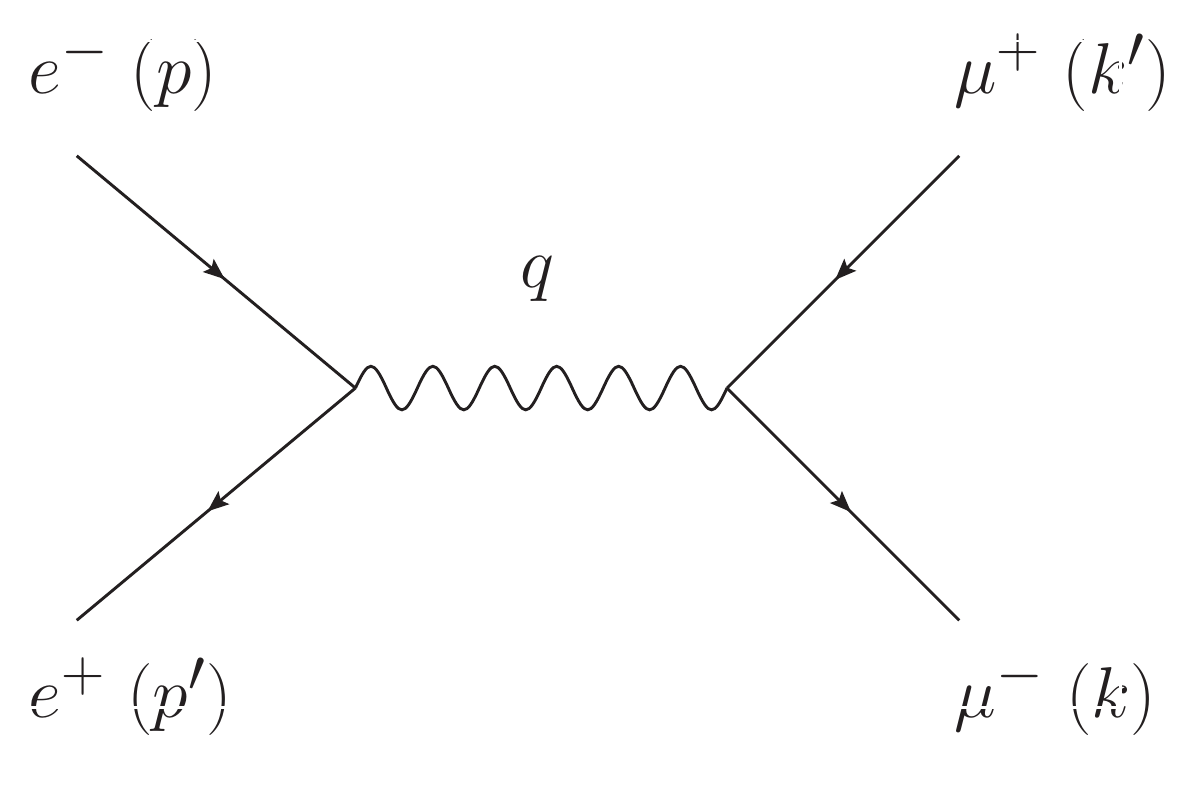}
  \caption{Feynman diagram for $e^+e^- \rightarrow \mu^+ \mu^-$ in QED via photon exchange.}
  \label{fig:eemumu}
\end{figure}

The steps for calculating the matrix element and hence differential cross section for this process are given, for example, in Ref.~\cite{Peskin:1995ev}, Ch. 5. Here we list the main steps in the calculation of $e^+e^- \rightarrow \mu^+ \mu^-$ in QED via photon exchange. The Feynman diagram for this process is shown in Fig.~\ref{fig:eemumu}. 
Using the QED Feynman rules, one can immediately write down the amplitude:
\begin{equation}
i \mathcal{M} = \bar{v}^{s'} (p') (-ie \gamma^\lambda) u^s(p) \left( \frac{ - i g_{\lambda\nu} } { q^2 }\right)\bar{u}^r (k) ( - ie \gamma^\nu ) v^{r'}(k') \;,
\end{equation}
where $s, s', r, r'$ are the spin indices. Writing them implicitly, the squared matrix element is given by
\begin{equation}
|\mathcal{M}|^2 = \frac{e^4}{q^4} ( \bar{v}(p') \gamma^\lambda u(p) \bar{u}(p) \gamma^\nu v(p'))  ( \bar{u}(k) \gamma_\lambda v(k') \bar{v}(k') \gamma_\nu u(k)) \;.
\end{equation}
For simplicity, we can average over the electron and positron spins and sum over the muon spins:
\begin{equation}
\frac{1}{2} \sum_s \frac{1}{2} \sum_{s'} \sum_r \sum_{r'} |\mathcal{M}|^2 \;.
\end{equation}
Using completeness relations for the spinors we can write:
\begin{equation}
\frac{1}{4} \sum_\mathrm{spins} = \frac{e^4} { 4 q^4 } \mathrm{Tr} [ \slashed{p}' \gamma^\lambda \slashed{p} \gamma^\nu ]\mathrm{Tr} [ \slashed{k} \gamma_\lambda \slashed{k}' \gamma_\nu ] \;,
\end{equation}
where we have neglected both the electron and muon masses. Using identities of traces of gamma matrices, one can show that:
\begin{equation}
\frac{1}{4} \sum_\mathrm{spins} = \frac{8 e^4} { q^4 } \left[ (p\cdot k)(p'\cdot k') + (p \cdot k' ) ( p' \cdot k) \right] \;.
\end{equation}

\begin{figure}[!htb]
  \centering
    \includegraphics[width=0.7\linewidth]{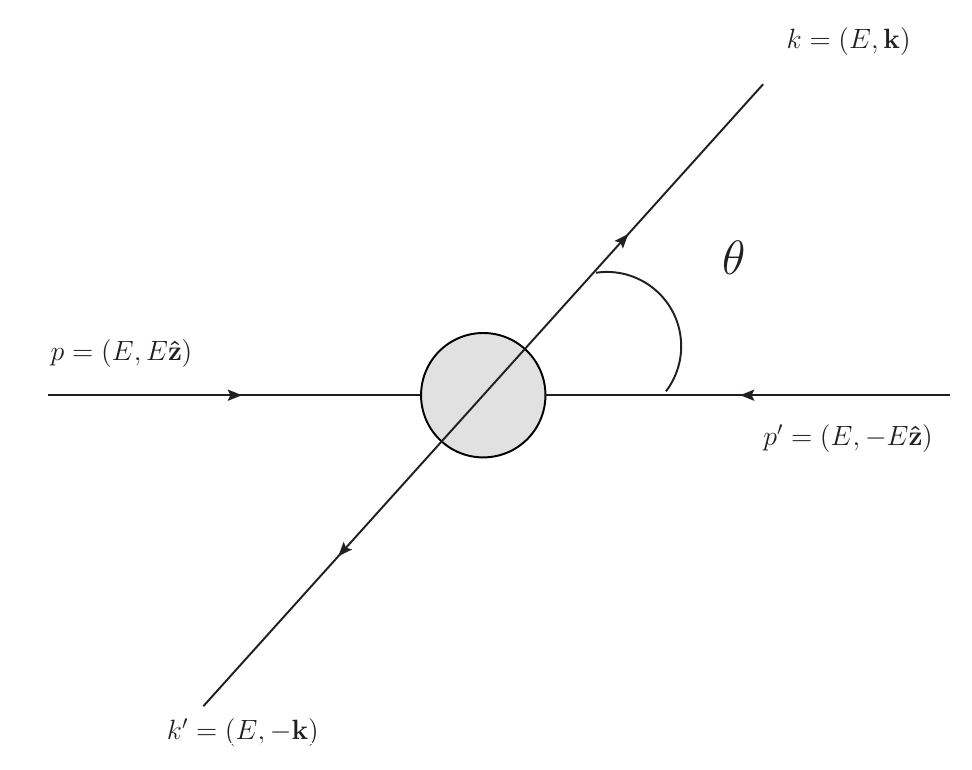}
  \caption{Schematic diagram for the kinematic setup of the process $e^+e^- \rightarrow \mu^+ \mu^-$. The angle $\theta$ is defined between the incoming electron and the outgoing muon, both being particles.}
  \label{fig:eemumukin}
\end{figure}

Up to this point the matrix element squared is expressed in terms of invariant dot products. To obtain a more explicit formula we must specialise to a particular frame of reference and write down expressions for the four-vectors of the particles involved in the collision. These are shown in Fig.~\ref{fig:eemumukin}. 
Using the four-vector explicit expressions we can express the invariants as:
\begin{eqnarray}
q^2 = (p+p') = 4 E^2&,&\;\; p\cdot p' = 2 E^2,\nonumber \\
p\cdot k = p' \cdot k' = E^2 - E |\mathbf{k}| \cos \theta&,&\;\; p\cdot k' = p' \cdot k = E^2 + E |\mathbf{k}| \cos \theta \;,
\end{eqnarray}
where the angle $\theta$ is defined in the figure. At high enough energies we can neglect the lepton masses, $E=|\mathbf{k}|$ and:
\begin{equation}
\frac{1}{4} \sum_\mathrm{spins} |\mathcal{M} |^2 =  e^4 ( 1 + \cos ^2 \theta ) \;.
\end{equation}
One can immediately plug the above expression into the relevant formula for the differential cross section for $2\rightarrow 2$ scattering: 
\begin{equation}
\frac{ \mathrm{d} \sigma } { \mathrm{d} \Omega } = \frac{1} { 2 E_\mathcal{A} 2 E_\mathcal{B }  | v_\mathcal{A} - v_\mathcal{B}| } \frac{ |\mathbf{k}| } { (2\pi)^2 4 E_\mathrm{cm} } |\mathcal{M} |^2\;,
\end{equation}
where $E_{\mathrm{cm}}$ is the centre of mass energy of the colliding particles, the difference $|v_\mathcal{A} - v_\mathcal{B}|$ is the relative velocity of the beams as viewed from the laboratory frame, $E_\mathcal{A}$, $E_\mathcal{B}$ their energies in that frame and $\mathrm{d} \Omega = \mathrm{d} \cos \theta~ \mathrm{d} \phi$ is the phase space factor. The result is then:
\begin{equation}\label{eq:qedxs}
\frac{ \mathrm{d} \sigma } { \mathrm{d} \Omega } = \frac{\alpha^2}{4 \hat{s} } ( 1  + \cos ^2 \theta ) \;,
\end{equation}
where $\alpha = e^2 / (4 \pi)$ is the QED running coupling. Since the expression does not depend on the angle $\phi$, we may integrate over it: this introduces a multiplicative factor of $2\pi$ on the RHS. We have also defined, $\hat{s} \equiv E_\mathrm{cm}^2$. 
\subsubsection{$e^+e^-\rightarrow  Z/\gamma \rightarrow \mu^+ \mu^-$} 
The differential cross section for electroweak production of $\mu^+\mu^-$ at a lepton collider proceeds in much the same way as the one in QED.  The main difference arises from the fact that the $Z$ boson couples with different strengths to left- and right-handed fermions~\cite{HalzenMartin}. Table~\ref{tb:couplings} shows the couplings of fermions to the $Z$ boson, in the form: 
\begin{equation}
\mathcal{L}_{ffZ} = - \frac{g_W}{2 \cos \theta_W} \sum_f \bar{\psi}_f \gamma^\mu (V_f - A_f \gamma_5) \psi_f Z_\mu\;,
\end{equation} 
where $g_W$ is the SU(2) coupling constant in the standard model, $\cos \theta_W$ is the cosine of the Weinberg angle, numerical values of which are found in Appendix~\ref{app:constants}, $\psi_f$ represents fermion $f$ and $Z_\mu$ is the $Z$ boson field strength. 
\begin{table}[!htb]
  \begin{center}
    \begin{tabularx}{\linewidth}{XXXX}
      \toprule
        fermions & $Q_f$ & $V_f$ & $A_f$ \\ \midrule
  u, c, t & $+\frac{2}{3}$ & $(+\frac{1}{2} - \frac{4}{3} \sin ^2 \theta_W )$ & $+\frac{1}{2}$ \\
  d,s, b & $-\frac{1}{3}$ & $(-\frac{1}{2} - \frac{2}{3} \sin ^2 \theta_W )$ & $-\frac{1}{2}$ \\
  $\nu_e$, $\nu_\mu$, $\nu_\tau$ & $0$ & $\frac{1}{2}$ & $+\frac{1}{2}$ \\
  $e$, $\mu$, $\tau$ & $-1$ & $(-\frac{1}{2} + 2 \sin ^2 \theta_W )$ & $-\frac{1}{2}$ \\\bottomrule
    \end{tabularx}
  \end{center}
  \caption{Couplings of fermions to the $Z$ boson, taken from Ref.~\cite{Ellis:1991qj}.}
\label{tb:couplings}
\end{table}
The difference is manifested in the resulting outgoing lepton distributions as an asymmetry between the forward and backward directions. While Eq.~\ref{eq:qedxs} contains only constant terms and terms proportional to the square of the cosine of the scattering angle, the inclusion of the $Z$ boson induces a term linear in $\cos \theta$: 
\begin{equation}\label{eq:partonicxs}
\frac{ \mathrm{d} \sigma } { \mathrm{d} \Omega } = \frac{\alpha^2}{4 \hat{s} } \left[ A_0 ( 1  + \cos ^2 \theta ) + A_1 \cos \theta \right] \;,
\end{equation}  
where $A_0$ and $A_1$ are given by:
\begin{eqnarray}
A_0 &=& Q_f^2 - 2 Q_f V_\mu  V_f ~\chi_1 + (A_\mu^2 + V_\mu^2)  (A_f^2 + V_f^2)  ~\chi_2\;, \nonumber \\
A_1 &=& -4 Q_f  A_\mu A_f ~\chi_1 + 8 A_\mu V_\mu A_f V_f ~ \chi_2\;,
\end{eqnarray}
where in turn, the functions $\chi_1$ and $\chi_2$ are given by:
\begin{eqnarray}
\chi_1 (\hat{s}) &=& \kappa \hat{s} ( \hat{s} - M_Z^2 ) / (  (\hat{s}-M_Z^2)^2 + \Gamma_Z^2 M_Z^2 ) \;, \nonumber \\
\chi_2 (\hat{s}) &=&  \kappa^2 \hat{s}^2 / (  (\hat{s}-M_Z^2)^2 + \Gamma_Z^2 M_Z^2 ) \;, \nonumber \\
\kappa &=& \sqrt{2} G_f M_Z^2 / (4 \pi \alpha) \;.
\end{eqnarray}
A good test to check whether the Monte Carlo integration is working is to check whether the Monte Carlo cross section agrees with the analytic result:
\begin{equation}\label{eq:sigmaee}
\sigma = \frac{4 \pi \alpha^2} { 3 \hat{s} } A_0 \;,
\end{equation}
where it is evident that the $\cos \theta$ term has dropped out due to its asymmetry. 
\subsection{Exercise 1: lepton colliders}
In this exercise the aim is to produce a Monte Carlo event generator for $e^+e^- \rightarrow  Z/\gamma \rightarrow \mu^+ \mu^-$. Of course the choice of `final' flavour is arbitrary, since we have neglected all lepton masses to this point. Note, however, that if one wants to consider $e^+e^- \rightarrow   e^+ e^-$, then there exists a new $t$-channel diagram that is not included in the above expression. 

The integration to obtain the cross section is in fact trivial, since we know how to integrate cosine functions analytically, and the $e^+e^-$ centre-of-mass energy, $\hat{s}$, is fixed, without requiring any Jacobian transformations to improve efficiency (i.e. there's no $\mathrm{d} m^2$ integral as in Eq.~\ref{eq:bwint}). Nevertheless, the exercise provides an insight to the basic building blocks of an event generator. The algorithm is given in Section~\ref{sec:hitmiss}. One thing to notice is that to obtain the cross sections in picobarn, one has to use the conversion factor in Table~\ref{tb:constants} in Appendix~\ref{app:constants}.

The example `solution' was written in \verb=Python=, and provides some basic plotting using \verb+Matplotlib+. A histogram of the only variable $\cos \theta$ is given. In this case this is an observable that we can measure, since we know both the direction of the incoming lepton and the outgoing lepton (between which this angle is defined). Moreover, the momenta are `set up' in the laboratory frame, which is equivalent to the centre-of-mass frame in this case. 

Some suggestions for possible extensions:
\begin{itemize}
\item Check the cross section against the analytical formula. For example, at $E_\mathrm{cm} = 90$~GeV: $\sigma = 1060.82 \pm0.25$~pb versus the analytic result: $\sigma_\mathrm{analytic} = 1060.93$~pb. 
\item Plot distributions of the energy of particles, or the pseudo-rapidity (in this case equal to the rapidity since we neglect the mass): $\eta = - \ln \tan (\theta / 2)$. 
\item Investigate the forward-backward asymmetry: $A_\mathrm{FB} \equiv (\sigma_F - \sigma_B) / (\sigma_F + \sigma_B)$, where $\sigma_{F,B}$ are the forward (right `hemisphere', $\theta \in (-\pi/2, +\pi/2)$) and backward (left `hemisphere', $\theta \in (\pi/2, +\pi)\cup (-\pi/2, -\pi)$) cross sections respectively. 
\end{itemize}
\subsection{Exercise 2: hadron colliders}
The previous exercise involved essentially a one-dimensional integral, over the angle $\theta$. For an electron-positron collision, this is always the case for a $2\rightarrow 2$ hard process. The next incremental complication arises for hard processes at hadron colliders. Since the hadrons are not elementary particles, we have to consider collisions between their constituent quark and gluons (partons), at high enough energies ($E\gg 1$~GeV). This results in the following considerations:
\begin{itemize}
\item The centre-of-mass energy of the colliding partons is not fixed, i.e. $\hat{s}$ is variable. Moreover, since the centre-of-mass frame and the laboratory frame (where observations are made) are not the same, the final-state particles need to be Lorentz-boosted from one frame to the other, in order to construct observable distributions. 
\item We need to consider the distribution of momenta of the colliding partons inside the protons as well as the different contributing quark flavours, characterised by the parton density functions. The parton density function for flavour $q$ for a quark (or gluon) carrying momentum fraction $x$ of the proton at momentum transfers $Q^2$ is denoted by $f_q(x,Q^2)$. This can be accessed via the \verb+LHAPDF+ library~\cite{Whalley:2005nh, lhapdfsite}. For more details on the \verb+LHAPDF+ interface, see Appendix~\ref{app:pdf}. 
\item Due to the above two points, we now have essentially four variables that characterise the phase space: $\hat{s}$, the momentum fractions $x_{1,2}$ and the scattering angle $\theta$, plus one constraint allowing us to eliminate one: $\hat{s} = x_1 x_2 S$, where $S$ is the proton-proton centre of mass energy squared. This leaves us with a 3-dimensional phase space for the hard process at hadron colliders. 
\item When summing over quark flavours, one has to note that the angle $\cos \theta$ is defined with respect to the incoming particle (as opposed to anti-particle) and the outgoing particle (as opposed to anti-particle). This implies that for example, in a collision of $u \bar{u}$, if $\theta$ is defined with respect to the positive $z$-axis, one must add a contribution for $\bar{u} u$, with $ \theta \rightarrow \pi - \theta$, resulting in the change $\cos \theta \rightarrow - \cos \theta$. Effectively this cancels out the asymmetric part of the distribution in a proton-proton collider (but not in a $p \bar{p}$ collider such as the Tevatron). 
\item For the purposes of this exercise we will cut-off the di-lepton invariant mass at some value, $Q_\mathrm{min}$. This will appear in the limits of the integrals we perform. For reasonable results, we will choose $Q_\mathrm{min} = 60$~GeV. 
\item The matrix element squared has to be multiplied by a factor of $1/3$: this \textit{averages} over the initial quark-anti-quark colour configurations. If we also had quarks in the final state, we would need to \textit{sum} over their colours. 
\end{itemize}
The partonic cross section of Eq.~\ref{eq:partonicxs} is still valid in the case of $q\bar{q} \rightarrow Z/\gamma \rightarrow \mu^+ \mu^-$, with the quark charges taken into consideration accordingly. However, we must now consider the hadronic cross section:
\begin{equation}
\frac{\mathrm{d} \sigma} { \mathrm{d} \hat{s} ~\mathrm{d} \cos \theta } = \sum_{q,q'}\int_0^1 \mathrm{d} x_1  \int_0^1\mathrm{d} x_2~ \delta ( \hat{s} - x_1 x_2 S  )~ f_q(x_1, \hat{s}) f_{q'}(x_2, \hat{s}) ~\frac{\mathrm{d} \hat{\sigma}} {\mathrm{d} \cos \theta } \;,
\end{equation}
with $\mathrm{d} \hat{\sigma}/\mathrm{d} \cos \theta$ given by Eq.~\ref{eq:partonicxs}, and we have already made the replacement $Q^2 = \hat{s}$ for the PDF factorisation scale. The sum is written here generically, over $q$ and $q'$ but should be taken over $q\bar{q}$ for the process we are considering. The integral over the $\delta$-function can then be performed to eliminate one of the dependent observables. We remove $x_2$ and remove the integral over $x_1$, turning it into a differential on the left-hand side:
\begin{eqnarray}
\frac{\mathrm{d} \sigma} { \mathrm{d} \hat{s} ~ \mathrm{d} x_1 ~\mathrm{d} \cos \theta } &=& \int_0^1\mathrm{d} x_2~ \delta ( S x_1 (x_2 - \hat{s}/(S x_1) )~ f_q(x_1, \hat{s}) f_{q'}(x_2, \hat{s}) ~\frac{\mathrm{d} \hat{\sigma}} {\mathrm{d} \cos \theta } \;, \nonumber \\
&=& \frac{1}{\hat{s} x_1}~ f_q(x_1, \hat{s}) f_{q'}(x_2=\hat{s}/(Sx_1), \hat{s}) ~\frac{\mathrm{d} \hat{\sigma}} {\mathrm{d} \cos \theta } \;. 
\end{eqnarray}
We define $\tau \equiv \hat{s} / S$ and the rapidity of the outgoing di-lepton system: 
\begin{equation}
y \equiv \frac{1}{2} \ln \left( \frac{E+p_z}  { E-p_z } \right) = \frac{1}{2} \ln \left( \frac{x_1} {x_2} \right) \;,
\end{equation}
by which $x_{1,2} = \sqrt{\tau} \mathrm{e}^{\pm y}$, and:
\begin{equation}
\mathrm{d} x_1 \mathrm{d} \hat{s}/(\hat{s} x_1) = \mathrm{d} \tau \mathrm{dy}\;.
\end{equation}
We finally arrive at: 
\begin{equation}
\frac{\mathrm{d} \sigma} { \mathrm{d} \tau ~ \mathrm{d} y ~\mathrm{d} \cos \theta }  = \sum_{q,q'} ~ f_q(x_1=\sqrt{\tau} \mathrm{e}^{+y}, \hat{s} = \tau S) f_{q'}(x_2= \sqrt{\tau} \mathrm{e}^{-y}, \hat{s}=\tau S) ~\frac{\mathrm{d} \hat{\sigma}} {\mathrm{d} \cos \theta }\;.
\end{equation}
The integration over the phase space can be performed via the Monte Carlo method by selecting $\tau$, $y$ and $\cos \theta$ randomly. Since we know we have a heavy resonance (the $Z$ boson) in the process, we can attempt to perform a Jacobian transformation as was described in Section~\ref{sec:jacob}. Note, however, that in this case the phase space is not flat after transformation since we have the photon contribution at low invariant masses, as well as the interference contribution. Nevertheless, the transformation is still useful and it is recommended. One can experiment with the parameters of the transformation relation to see if the variance can be decreased by clever choices. Hence, for random numbers $R_i \in (0,1)$, $i = 1,2$: 
\begin{eqnarray}
\cos \theta &=& 2 R _1 - 1 \nonumber \\
y &=&   (2 R_2 - 1 ) y_\mathrm{max}   \;, 
\end{eqnarray}
with the maximum value of the rapidity given by: $y_\mathrm{max} = - 0.5 \ln (\tau)$. 
The $\tau$-integral has to be more carefully considered. Defining the transform mass and width parameters $M_\mathrm{tr}$ and $\Gamma_\mathrm{tr}$, respectively, and keeping them free for the moment, we have:
\begin{equation}
\tau S = \hat{s} = M_\mathrm{tr} \Gamma_{tr} \tan (\rho) + M_{\mathrm{tr}}^2 \nonumber \\
\end{equation}
with $\rho$ in $(\rho_\mathrm{min}, \rho_\mathrm{max})$, generated using random number $R_3 \in (0,1)$ via:
\begin{equation}
\rho = \rho_{\mathrm{min}} + (\rho_\mathrm{max} - \rho_\mathrm{min}) R_3 \;,
\end{equation}
where $\rho$ is limited by the choice of $Q_\mathrm{min}$ and the hadron centre-of-mass energy $\sqrt{S}$: 
\begin{eqnarray}
\rho_\mathrm{min} = \tan^{-1} \left( \frac{Q_\mathrm{min}^2 - M_\mathrm{tr}^2} {\Gamma_\mathrm{tr} M_\mathrm{tr}}\right) \;, \nonumber \\ 
\rho_\mathrm{max} = \tan^{-1} \left( \frac{S - M_\mathrm{tr}^2} {\Gamma_\mathrm{tr} M_\mathrm{tr}}\right)\;.
\end{eqnarray}
The integration can be performed in an equivalent way as in Exercise 1, and the maximum weight can be stored to perform the `hit-or-miss' unweighing of events. This is again, exactly equivalent to the case of lepton colliders. A final complication for the case of the hard processes at hadron colliders is boosting between the centre-of-mass frame (where the calculation of the partonic cross section was performed) into the lab frame. We already know the 4-momenta in the lab frame for the incoming partons:
\begin{eqnarray}
p_q^\mathrm{lab} &=& \frac{\sqrt{\hat{s}}}{2} (x_1, 0, 0, x_1) \; \nonumber \\
p_{q'}^\mathrm{lab} &=& \frac{\sqrt{\hat{s}}}{2} (x_2, 0, 0, -x_2)  \;,
\end{eqnarray}
where $\sqrt{\hat{s}} = E_\mathrm{cm}$ is the centre-of-mass frame collision energy of the partons. The Lorentz boost factor along the $z$-axis between the lab and centre-of-mass frames can be calculated and is given by:
\begin{equation}
\beta = \frac{x_2 - x_1} {x_2 + x_1}\;,
\end{equation}
where $\beta = v/c$. And hence, the momenta in the centre-of-mass frame:
\begin{eqnarray}
p_\mu^\mathrm{cm} &=& \frac{\sqrt{\hat{s}}}{2} (1,\sin \theta \cos \phi, \sin \theta \sin \phi, \cos \theta) \; \nonumber \\
p_{\bar{\mu}}^\mathrm{cm} &=& \frac{\sqrt{\hat{s}}}{2} (1,-\sin \theta \cos \phi, -\sin \theta \sin \phi,- \cos \theta) \;,
\end{eqnarray}
(where $\phi$ has been generated randomly and uniformly using a random number $R_4 \in (0,1)$: $\phi = 2 \pi R_4$) can be transformed into those in the lab frame via a Lorentz boost along the $z$-direction:
\begin{equation}
p^\mathrm{lab} = ( \gamma p_0 - \gamma \beta p_3, p_1, p_2, -\gamma \beta p_0 + \gamma p_3 )\;,
\end{equation}
where $\gamma = \sqrt( 1  / (1 - \beta^2) )$. 

The solution to this exercise is provided as a \verb+Python+ program as well, and generates a set of histograms using the \verb+Matplotlib+ library.

Some suggestions for further investigations:
\begin{itemize}
\item Calculate the cross sections for di-lepton production via $Z\gamma$ at proton-proton colliders 8~TeV and 14~TeV using the \verb+cteq6l1+ PDF sets and compare to the \texttt{MadGraph} results: $\sigma(8~\mathrm{TeV})=(881.8\pm 1)$~pb and $\sigma(14~\mathrm{TeV})=(1684\pm 1.3)$~pb. Note that the minimum same-flavour lepton invariant mass was taken to be $60$~GeV and no other cuts were imposed on the leptons. 
\item Consider the modifications necessary to simulate a $p\bar{p}$ collider. 
\item The Les Houches file format allows one to write parton-level events and feed them into a general-purpose Monte Carlo for parton showering and hadronization. An explanation of how the format looks like is found in Appendix~\ref{app:fileformat}. 
\end{itemize}

\section{After the hard process}
Even though we will not go into the technical details of the implementation of the following steps in event generation, it is interesting to list some of the considerations necessary to perform them. The factorised view of Monte Carlo event generation has already been illustrated by Figs.~\ref{fig:step1} to~\ref{fig:step5}. Step-by-step, some points that need to be considered are:
\begin{enumerate}
\item The hard process can be $2\rightarrow N$, where $N$ is any number of particles. 
\item Decays can be easily implemented on top of any process in a factorised way, given that the resonance is narrow enough. If this is the case, one can consider the decay of a massive resonance in its rest frame, and then boost the decay products into the lab frame according to the particle's boost in that frame. 
\item Most parton showers are based on collinear and soft splitting kernels that capture the enhanced regions. There are two possibilities for parton showers, with some technical differences in the implementation: radiation from final-state particles or radiation from initial-particles. The difference arises because initial-state particles need to `evolve' back to the incoming hadrons, whereas final-state particles have to evolve forward to hadrons. 
\item At some scale, $\mathcal{O}(1~\mathrm{GeV})$, perturbation theory breaks down and a non-perturbative model needs to take over. The phenomenon is called hadronization. The outgoing quarks and gluons need to be treated through some model that groups them into QCD colour-singlets. This is done in \texttt{HERWIG 7}, for example, via a cluster model, and in \texttt{Pythia 8} via a string model. Another non-perturbative effect involves the interaction of multiple partons. In \texttt{HERWIG 7} and \texttt{Sherpa} this phenomenon modelled as multiple QCD $2\rightarrow 2$ interactions~\cite{Bahr:2008dy,Gleisberg:2008ta}. In \texttt{Pythia 8} it is treated as being formed through interleaved parton-parton interactions in a common sequence with the initial-state radiation~\cite{Sjostrand:2004ef}.
\end{enumerate}
\section{Conclusions}
We have presented a short introduction to Monte Carlo event generators and directly delved into two simple examples. Solutions to the exercises are given and motivations on how one can go beyond were presented. 

\section{Acknowledgements}
The author would like to thank Christoph Grab and Nicolas Chanon the opportunity to lecture at the ``Advanced Scientific Computing Workshop'' at ETH Z\"urich, as well the students who attended the course, providing helpful feedback. Support is acknowledged in part by the Swiss National Science Foundation (SNF) under contract 200020-149517, by the European Commission through the ``LHCPhenoNet'' Initial Training Network PITN-GA-2010-264564, MCnetITN FP7 Marie Curie Initial Training Network PITN-GA-2012-315877 and by a Marie Curie Intra European Fellowship within the 7th European Community Framework Programme (grant no. PIEF-GA-2013-622071). 
\bibliographystyle{unsrtnat}
\appendix 

\section{Constants}\label{app:constants}
The constants in this section are given to provide agreement with the \texttt{MadGraph} event generator. They appear in Table~\ref{tb:constants}.
\begin{table}[!htb]
  \begin{center}
    \begin{tabularx}{\linewidth}{XXX}
      \toprule
        variable & symbol & value \\ \midrule
        conversion factor & GeV$^{-2}\rightarrow$~pb & $3.894\times 10^8$~pb per GeV$^{-2}$\\
        $Z$ boson mass & $M_Z$ & 91.188~GeV \\
        $Z$ boson width & $\Gamma_Z$ & 2.4414~GeV \\
QED running coupling & $\alpha$ & $\frac{1}{132.507}$ \\
Fermi constant & $G_f$ & $1.16639\times 10^{-5}$~GeV$^{-2}$. \\
Weinberg angle & $\sin^2 \theta_W$  & 0.222246  \\\bottomrule
    \end{tabularx}
  \end{center}
  \caption{Constants used throughout this article, given to provide agreement with \texttt{MadGraph}.}
\label{tb:constants}
\end{table}
\section{Parton density functions using LHAPDF}\label{app:pdf}
At the time of writing, the latest version of the \verb+LHAPDF+ package is 6.1.3. It is recommended to use this or a latter version for the exercises given here. The library can be interfaced to either \verb=C++=, \verb+FORTRAN+ or \verb+Python+. Since the solutions to the exercises are given in \verb+Python+, the PDFs should be initialised as:
\begin{verbatim}
## import LHAPDF and initialise PDFs
import lhapdf
## initialises PDF member object (for protons)
p = lhapdf.mkPDF("cteq6l1", 0)
\end{verbatim}
and the PDF should be called as:
\begin{verbatim}
p.xfxQ(FLAVOUR, x1, mu)
\end{verbatim}
where \verb+FLAVOUR+ should be replaced by the quark flavours contributing to the process: 1 for down-quarks, 2 for up, 3 for strange, 4 for charm and negative values for the corresponding anti-quarks. The gluon, not used here, is given by 21. Note that this actually gives $x \times f(x)$ and thus one has to divide by the momentum fraction to get $f(x)$. Moreover, this specific function takes as input the scale and \textit{not} the scale squared. 

\section{The Les Houches event file format}\label{app:fileformat}
The file header and the first event in a Les Houches-accord event file have the following form:
{\tiny
\begin{verbatim}
<LesHouchesEvents version="1.0">

<header>

...

</header>
<init>
     2212     2212  0.40000000000E+04  0.40000000000E+04 0 0   10042   10042 2   1
  0.88184317905E+03  0.10037036184E+01  0.86172440000E-01   0
</init>
<event>
 5   0  0.4467596E-01  0.9118800E+02  0.7546771E-02  0.1300000E+00
       -2   -1    0    0    0  501  0.00000000000E+00  0.00000000000E+00  0.10230021267E+01  0.10230021267E+01  0.00000000000E+00 0.  1.
        2   -1    0    0  501    0  0.00000000000E+00  0.00000000000E+00 -0.21100317982E+04  0.21100317982E+04  0.00000000000E+00 0. -1.
       23    2    1    2    0    0  0.00000000000E+00  0.00000000000E+00 -0.21090087961E+04  0.21110548003E+04  0.92920762309E+02 0.  0.
      -11    1    3    3    0    0  0.42119725672E+01 -0.21951919980E+02 -0.12916294295E+03  0.13108277284E+03  0.00000000000E+00 0.  1.
       11    1    3    3    0    0 -0.42119725672E+01  0.21951919980E+02 -0.19798458531E+04  0.19799720275E+04  0.00000000000E+00 0. -1.
</event>

...


</LesHouchesEvents>
\end{verbatim} 
}
The first row after \verb+<init>+  shows the ids of the incoming hadrons, their energy and the PDF numbers (10042 in this case). The following line shows the cross section and the error. The first event follows, containing the particle ids, their status codes and mother information, colour information, their momenta, and whether they are stable particles or not. See Ref.~\cite{Alwall:2006yp} for more details. 

\section{Convergence}\label{app:convergence}

\begin{table}[!htb]
  \begin{center}
    \begin{tabularx}{\linewidth}{XXX}
      \toprule
Technique & Convergence & worse than MC in d > \\ \midrule
trapezium & $1/N^{2/d}$ & 4 \\ 
Simpson's & $1/N^{4/d}$ & 8 \\ 
$m$th-order gaussian quadrature & $1/N^{(2m-1)/d}$ & $4m - 2$ \\
Monte Carlo & $1 / \sqrt{N}$ & -\\ \bottomrule
    \end{tabularx}
  \end{center}
\caption{The rate of convergence with the number of points $N$ used for
  each method in $d$-dimensions.}
\label{tb:convergence}
\end{table}

\bibliography{mchowto.bib}

\begin{thebibliography}{26}
\providecommand{\natexlab}[1]{#1}
\providecommand{\url}[1]{\texttt{#1}}
\expandafter\ifx\csname urlstyle\endcsname\relax
  \providecommand{\doi}[1]{doi: #1}\else
  \providecommand{\doi}{doi: \begingroup \urlstyle{rm}\Url}\fi

\bibitem[Wikipedia(2020)]{wikipediahowto}
Wikipedia.
\newblock {How-to}.
\newblock https://en.wikipedia.org/wiki/How-to, 2020.

\bibitem[Twain(1897)]{twain}
M.~Twain.
\newblock \emph{{Following the Equator}}.
\newblock Sun-Times Media Group, 1897.

\bibitem[Alwall et~al.(2011)Alwall, Herquet, Maltoni, Mattelaer, and
  Stelzer]{Alwall:2011uj}
Johan Alwall, Michel Herquet, Fabio Maltoni, Olivier Mattelaer, and Tim
  Stelzer.
\newblock {MadGraph 5 : Going Beyond}.
\newblock \emph{JHEP}, 06:\penalty0 128, 2011.
\newblock \doi{10.1007/JHEP06(2011)128}.

\bibitem[Alwall et~al.(2014)Alwall, Frederix, Frixione, Hirschi, Maltoni,
  Mattelaer, Shao, Stelzer, Torrielli, and Zaro]{Alwall:2014hca}
J.~Alwall, R.~Frederix, S.~Frixione, V.~Hirschi, F.~Maltoni, O.~Mattelaer,
  H.~S. Shao, T.~Stelzer, P.~Torrielli, and M.~Zaro.
\newblock {The automated computation of tree-level and next-to-leading order
  differential cross sections, and their matching to parton shower
  simulations}.
\newblock \emph{JHEP}, 07:\penalty0 079, 2014.
\newblock \doi{10.1007/JHEP07(2014)079}.

\bibitem[Bahr et~al.(2008{\natexlab{a}})]{Bahr:2008pv}
M.~Bahr et~al.
\newblock {Herwig++ Physics and Manual}.
\newblock \emph{Eur. Phys. J.}, C58:\penalty0 639--707, 2008{\natexlab{a}}.
\newblock \doi{10.1140/epjc/s10052-008-0798-9}.

\bibitem[Arnold et~al.(2012)]{Arnold:2012fq}
K.~Arnold et~al.
\newblock {Herwig++ 2.6 Release Note}.
\newblock 2012.

\bibitem[Gieseke et~al.(2011)]{Gieseke:2011na}
S.~Gieseke et~al.
\newblock {Herwig++ 2.5 Release Note}.
\newblock 2011.

\bibitem[Bellm et~al.(2019)]{Bellm:2019zci}
Johannes Bellm et~al.
\newblock {Herwig 7.2 Release Note}.
\newblock 2019.

\bibitem[Bellm et~al.(2017)]{Bellm:2017bvx}
Johannes Bellm et~al.
\newblock {Herwig 7.1 Release Note}.
\newblock 2017.

\bibitem[Bellm et~al.(2016)]{Bellm:2015jjp}
Johannes Bellm et~al.
\newblock {Herwig 7.0/Herwig++ 3.0 release note}.
\newblock \emph{Eur. Phys. J.}, C76\penalty0 (4):\penalty0 196, 2016.
\newblock \doi{10.1140/epjc/s10052-016-4018-8}.

\bibitem[Bellm et~al.(2013)]{Bellm:2013hwb}
J.~Bellm et~al.
\newblock {Herwig++ 2.7 Release Note}.
\newblock 2013.

\bibitem[Sjostrand et~al.(2006)Sjostrand, Mrenna, and Skands]{Sjostrand:2006za}
Torbjorn Sjostrand, Stephen Mrenna, and Peter~Z. Skands.
\newblock {PYTHIA 6.4 Physics and Manual}.
\newblock \emph{JHEP}, 05:\penalty0 026, 2006.
\newblock \doi{10.1088/1126-6708/2006/05/026}.

\bibitem[Sjostrand et~al.(2008)Sjostrand, Mrenna, and Skands]{Sjostrand:2007gs}
Torbjorn Sjostrand, Stephen Mrenna, and Peter~Z. Skands.
\newblock {A Brief Introduction to PYTHIA 8.1}.
\newblock \emph{Comput. Phys. Commun.}, 178:\penalty0 852--867, 2008.
\newblock \doi{10.1016/j.cpc.2008.01.036}.

\bibitem[Gleisberg et~al.(2009)Gleisberg, Hoeche, Krauss, Schonherr, Schumann,
  Siegert, and Winter]{Gleisberg:2008ta}
T.~Gleisberg, Stefan. Hoeche, F.~Krauss, M.~Schonherr, S.~Schumann, F.~Siegert,
  and J.~Winter.
\newblock {Event generation with SHERPA 1.1}.
\newblock \emph{JHEP}, 02:\penalty0 007, 2009.
\newblock \doi{10.1088/1126-6708/2009/02/007}.

\bibitem[Buckley et~al.(2011)]{Buckley:2011ms}
Andy Buckley et~al.
\newblock {General-purpose event generators for LHC physics}.
\newblock \emph{Phys. Rept.}, 504:\penalty0 145--233, 2011.
\newblock \doi{10.1016/j.physrep.2011.03.005}.

\bibitem[Kittel et~al.(1970)Kittel, Van~Hove, and Wojcik]{Kittel:1970rh}
W.~Kittel, L.~Van~Hove, and W.~Wojcik.
\newblock {A monte carlo generation method with importance sampling for high
  energy collisions of hadrons}.
\newblock \emph{Comput. Phys. Commun.}, 1:\penalty0 425--436, 1970.
\newblock \doi{10.1016/0010-4655(70)90016-0}.

\bibitem[Lepage(1980)]{Lepage:1980dq}
G.Peter Lepage.
\newblock {VEGAS: AN ADAPTIVE MULTIDIMENSIONAL INTEGRATION PROGRAM}.
\newblock 3 1980.

\bibitem[Papaefstathiou(2011)]{Papaefstathiou:2011rc}
Andreas Papaefstathiou.
\newblock \emph{{Phenomenological aspects of new physics at high energy hadron
  colliders}}.
\newblock PhD thesis, Cambridge U., 2011.
\newblock URL \url{https://www.repository.cam.ac.uk/handle/1810/239399}.

\bibitem[Peskin and Schroeder(1995)]{Peskin:1995ev}
Michael~E. Peskin and Daniel~V. Schroeder.
\newblock \emph{{An Introduction to quantum field theory}}.
\newblock Addison-Wesley, Reading, USA, 1995.
\newblock ISBN 9780201503975, 0201503972.
\newblock URL \url{http://www.slac.stanford.edu/~mpeskin/QFT.html}.

\bibitem[Halzen and Martin(1984)]{HalzenMartin}
F.~Halzen and A.~Martin.
\newblock \emph{{Quarks and Leptons: An Introductory Course in Modern Particle
  Physics}}.
\newblock John Wilie and Sons, 1984.

\bibitem[Ellis et~al.(1996)Ellis, Stirling, and Webber]{Ellis:1991qj}
R.~Keith Ellis, W.~James Stirling, and B.~R. Webber.
\newblock {QCD and collider physics}.
\newblock \emph{Camb. Monogr. Part. Phys. Nucl. Phys. Cosmol.}, 8:\penalty0
  1--435, 1996.

\bibitem[Whalley et~al.(2005)Whalley, Bourilkov, and Group]{Whalley:2005nh}
M.~R. Whalley, D.~Bourilkov, and R.~C. Group.
\newblock {The Les Houches accord PDFs (LHAPDF) and LHAGLUE}.
\newblock In \emph{{HERA and the LHC: A Workshop on the implications of HERA
  for LHC physics. Proceedings, Part B}}, pages 575--581, 2005.

\bibitem[LHAPDF(2020)]{lhapdfsite}
LHAPDF.
\newblock http://lhapdf.hepforge.org/, 2020.

\bibitem[Bahr et~al.(2008{\natexlab{b}})Bahr, Gieseke, and
  Seymour]{Bahr:2008dy}
Manuel Bahr, Stefan Gieseke, and Michael~H. Seymour.
\newblock {Simulation of multiple partonic interactions in Herwig++}.
\newblock \emph{JHEP}, 07:\penalty0 076, 2008{\natexlab{b}}.
\newblock \doi{10.1088/1126-6708/2008/07/076}.

\bibitem[Sjostrand and Skands(2005)]{Sjostrand:2004ef}
T.~Sjostrand and Peter~Z. Skands.
\newblock {Transverse-momentum-ordered showers and interleaved multiple
  interactions}.
\newblock \emph{Eur. Phys. J. C}, 39:\penalty0 129--154, 2005.
\newblock \doi{10.1140/epjc/s2004-02084-y}.

\bibitem[Alwall et~al.(2007)]{Alwall:2006yp}
Johan Alwall et~al.
\newblock {A Standard format for Les Houches event files}.
\newblock \emph{Comput. Phys. Commun.}, 176:\penalty0 300--304, 2007.
\newblock \doi{10.1016/j.cpc.2006.11.010}.

\end{thebibliography}

\end{document}